\documentstyle[11pt,newpasp,twoside]{article}
\begin{document}

\title{Comparison of subclustering frequency determined from 2D and 3D data}
\author{Piotr Flin}

\affil{Joint Institute for Nuclear Physics, Bogoliubov Laboratory
of Theoretical Physics, Dubna, Russia\\
Pedagogical University, Institute of Physics,
ul. Swietokrzyska 15, 25--406, Kielce, Poland}

\author{Janusz Krywult}

\affil{Pedagogical University, Institute of Physics,
ul. Swietokrzyska 15, 25--406, Kielce, Poland}

\begin{abstract}
Frequencies of substructure within Abell clusters determined for the same
objects using the wavelet analysis for 2-D and 3-D data are similar.
\end{abstract}


The frequency of substructure occurrence has been determined by several
authors, using both various statistical methods and observational data.
Sometimes only the positions of galaxies, i.e. 2-D data, have been used, while
some papers have been based on the position and redshifts of member galaxies,
i.e. 3-D data. The determined frequency changed from below 30\% up to 80\% of
that for Abell clusters with substructures.

The analysis of previous papers suggests that the differences in the
determined substructure occurrences are due to the rather different methods of
analysis than to the observational data themselves. Therefore, in order to
test this hypothesis, in the present paper  a direct comparison of the
frequency of substructure occurrence is made for 2-D data and 3-D data for the
same Abell clusters.

All 43 galaxy clusters in the present work are the result of the objective
algorithms for star/galaxy classification applied to scanned photographic
plates.  There are three sources of our data, but the photographic material
came from 48" Schmidt telescopes. Each galaxy within the radius of 1.5 Mpc
($h=0.75$) from the cluster centre and with magnitude between $m_3$ and
$m_3+3$, where $m_3$ is the magnitude of the third brightest galaxy, was
considered as the cluster member. 33 catalogues of  Abell clusters were
obtained from DSS using FOCAS package, further 4 catalogues were obtained from
scans performed in the Rome Observatory (Trevese et al. 1992; Flin et al.
2000), while the remaining 6 catalogues of clusters came from the COSMOS
machine at the Royal Observatory in Edinburgh  (Krywult, MacGillivray, \& Flin
1999).

The existence of the substructures in galaxy clusters has been checked using
wavelet analysis. The wavelet technique is a convolution, on a grid of
$N\times N$ pixels, between signal $s(r)$ (in our case, the angular positions
of galaxies) and analyzing wavelet function $g(r,a)$. In this work we use
two-dimensional radial function called the Mexican Hat (Escalera et al. 1992).

For the analysis presented here, the discrete wavelet was computed on a grid
of $256\times 256$ pixels for seven scales increasing from $a=8$ to 64 (in
pixel units), namely $8, 11, 16, 22, 32, 45, 64$ respectively, which ensure
the correctness of the Mexican Hat (Daubechies 1990).

We have modelled the significance of the substructuring detected using the
Monte Carlo simulations. For each cluster and each scale $a$, the wavelet
analysis was carried out on a set of 1000 structureless distributions of
galaxies containing the same number of points as in the true fields.

We assume that a substructure is real if the probability of random
fluctuations is less than 0.005. Furthermore, for each scale $a$ only
substructures with more than 4 galaxy members in a circle of radius $a$ are
considered.

In order to compare our 2-D data with 3-D data, the Girardi et al. (1997)
paper served as a source of information on the subclustering in 3-D data.
They used wavelet analysis and performed morphological classification of
clusters (Escalera et al. 1994), finding bimodal structure for five clusters
(A548, A754, A1736, A3526, A3716) and a complex structure in A2151. Moreover,
the existence of substructure was detected in A85, A193, A194, A1060, A1367,
A1983, A2063, A2877 and A3667.

In our analysis, substructures were detected in clusters: A151, A426, A548,
A754, A1060, A1367, A 1736, A1809, A1983, A2052, A2151, A3128, A3395, A3526
and A3667.

Comparison shows that in some clusters with no subclustering in 2-D, the
analysis of 3-D data revealed small substructures, both in size and in the
number of members (A85, A193, A194, A2063). In a few cases, the 2-D data
showed substructures non-existing in the 3-D data (A151, A1809, A2052).

The differences in the substructure existence in the 2-D and the 3-D data do
occurr in individual cases. But the frequency of substructure occurrence in
3-D is 31\%, while our result, based on the 2-D data, is 34\%. At the
significance level of 0.05, these numbers are in agreement. So, the frequency
of the existence of substructures in Abell clusters determined from 2-D data
is a good indicator of subclustering frequency, which allows us to use it for
clusters located far away, when the number of galaxies with known redshifts is
small.

\noindent
This work was partially supported by grants KBN/IBM~SP/WSPKielce/073
and KBN/AS/BS/052.

\end{document}